\documentclass[
preprint
,aps
,secnumarabic
,amssymb, amsmath,nobibnotes, aps, prl, floatfix
]{revtex4-1}

\usepackage{wasysym}
\usepackage{afterpage}
\usepackage[RGB]{xcolor}
\usepackage{pdfcomment}
\usepackage{graphicx}
\usepackage{dcolumn}
\usepackage{bm}
\usepackage{color}
\usepackage{array}
\usepackage{float}
\usepackage{rotating}
\usepackage{setspace}
\usepackage{enumitem}
\setlist{nolistsep,leftmargin=*}
\usepackage{tikz}

\newcolumntype{P}[1]{>{\centering\arraybackslash}p{#1}}
\newcolumntype{M}[1]{>{\centering\arraybackslash}m{#1}}

\newcommand{\cmt}[1]{}

\begin{document}

\title{Surface plastic flow in polishing of rough surfaces}
\author{Ashif S. Iquebal}
\author{Dinakar Sagapuram}
\author{Satish Bukkapatnam\footnote{\label{Corresponding author}Corresponding author}}

\affiliation{
\fontsize{8}{12}{Department of Industrial and Systems Engineering, Texas A\&M University, College Station, Texas 77840, USA}\\
\text{\normalfont \normalsize ashif\_22@tamu.edu, dinakar@tamu.edu, satish@tamu.edu}
}

\begin{abstract}
We report electron microscopy observations of the surface plastic flow in polishing of rough metal surfaces with a controlled spherical asperity structure. We show that asperity--abrasive sliding contacts exhibit viscous behavior, where the material flows in the form of thin fluid-like layers. Subsequent bridging of these layers among neighboring asperities result in progressive surface smoothening. Our study provides new phenomenological insights into the long-debated mechanism of polishing. The observations are of broad relevance in tribology and materials processing. 

\end{abstract}

\maketitle

Mechanical interactions between severely rubbing surfaces have long been of fundamental interest for understanding friction in a wide range of domains including tribology, materials processing and geophysics. An important practical application of such interactions is in polishing of materials where rubbing action of fine abrasives is utilized to obtain smooth surfaces for application in optics, microscopy and mechanical instrumentation.

The practice of polishing to impart solid surfaces with smooth, lustrous finish has been known for centuries. The use of hard abrasives such as corundum and diamond for polishing in fact dates back to the Neolithic period \cite{lu2005earliest} and Leonardo da Vinci is credited with the earliest systematic design of a polishing machine \cite{pedretti1978codex}. It might be surprising then to know that the mechanism of polishing---how surface irregularities are smoothened out by abrasive particles---is still unsettled. Excellent account of the history and theories of polishing can be found in \cite{cornish1961mechanism,rabinowicz1968polishing,archard1985mechanical}. However, it may suffice to note that mainly two lines of thought for the polishing mechanism have prevailed: that of abrasion and surface flow. Early theories by Hooke and Newton \cite{newton1979opticks}, followed by those of  Herschel \cite{archard1985mechanical} and Rayleigh \cite{bulsara1998mechanics} viewed polishing essentially as an abrasion or a grinding process at a very fine scale where surface irregularities are removed by cutting action of the abrasives. The work by Samuels \cite{samuels2003metallographic} presented irrefutable evidence for this mechanism and showed how abrasives act as planing tools and result in the generation of well-defined chips as they slide past a surface. The alternative theory emerges from the work by Beilby \cite{beilby1921aggregation} who proposed surface smoothening occurring via surface flow and material redistribution. Here, it is believed that the material from surface peaks `flows' to fill up the valleys and forms a thin vitreous surface layer, generally referred to as the ``Beilby layer''. Bowden and Hughes \cite{bowden1937physical} further developed this theory and proposed that surface flow is in fact mediated by local melting at the surface--abrasive contacts. Electron diffraction measurements of polished surfaces have been presented as indirect evidence for the Beilby layer formation, but these observations were later proved to be inconclusive. To our knowledge, no conclusive evidence for the surface flow or melting has been provided to date. Other theories of polishing also exist, among which noteworthy is the molecular level material removal mechanism put forward by Rabinowicz \cite{rabinowicz1968polishing} based on energy considerations.

\begin{figure}
	\includegraphics[width=0.8\textwidth]{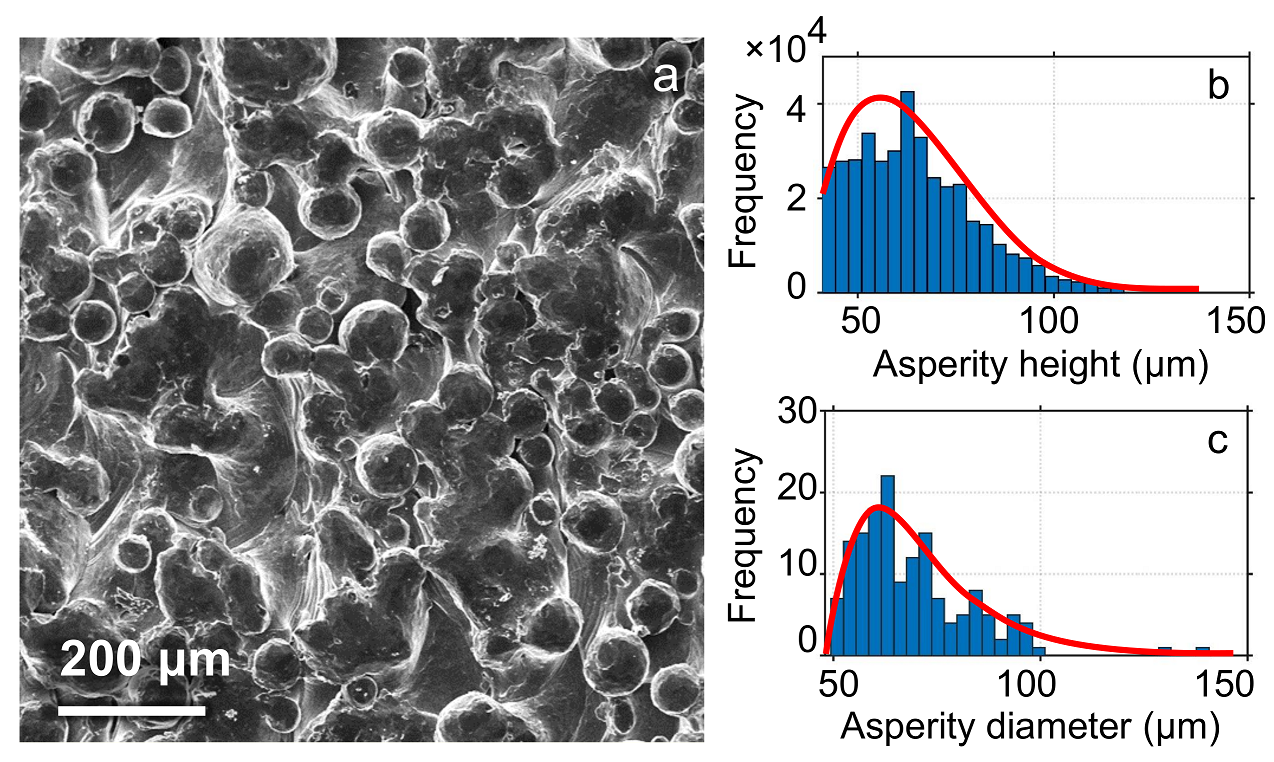}
	\centering
	\caption{Surface morphology characteristics of Ti-6Al-4V alloy sample prepared using electron beam melting. (a) Scanning electron micrograph showing the spherical asperity structure of the sample surface; (b) and (c) are respectively the distribution plots for the asperity height and diameter as measured using white light interferometry.}
	\label{fig:im1}
\end{figure}

In this study, we report direct experimental observations of surface plastic flow in polishing of an idealized rough metal surface having spherical asperities. Our electron microscopy observations of polished surfaces reveal viscous flow at the asperity--abrasive sliding contacts, involving surface material flow towards the asperity sides in the form of thin fluid-like layers. The subsequent stages of polishing involve bridging of these layers among different asperities to result in a smooth finish. Our study, besides confirming many hypotheses of the surface flow theory, provides new phenomenological insights into various stages of surface plastic flow in polishing of rough surfaces. 

Ti-6Al-4V samples of $\diameter$50 mm and 7 mm thickness with controlled surface topography consisting of spherical asperity structure were prepared using the electron beam melting process. Details of the processing conditions for generating this asperity structure are provided in the Supplemental Material. Scanning electron micrograph of the representative surface morphology is shown in Fig.~\ref{fig:im1}(a). The distributions of asperity height and diameter are shown in Figs.~\ref{fig:im1}(b) and \ref{fig:im1}(c), respectively. The asperity height as well as the diameter exhibits a Weibull distribution with an average value of 72 $\mu$m and 64.5 $\mu$m, respectively and a standard deviation of $\sim15$ $\mu$m. It may be noted that the idealization of surfaces as a collection of spherical asperities (with Gaussian and Weibull distribution of heights) has been the basis for many prior theoretical analyses of elastic--plastic contacts between rough surfaces \cite{Greenwood300,whitehouse1970properties,hutchings1992tribology}. 

{\color{black}The disk samples were polished on a Buehler Metaserv Grinder-Polisher (model 95-C2348-160) using silicon carbide (SiC) polishing pads ($\diameter$203 mm), in stages, with progressively smaller abrasives ranging from 30 $\mu$m to 5 $\mu$m under dry conditions. A steady nominal down pressure of $\sim0.5$ kPa was maintained and the polisher speed was fixed at 500 rpm. The workpiece sample was manually subject to a quasi-random orbital motion. The final polishing step involved the use of alumina abrasives ($< 1$ $\mu$m), suspended in an aqueous solution (20\% by wt., pH $\approx7.5$) for 20 minutes to impart a specular finish to the surface.} The polishing was interrupted at every 90 s intervals to observe the surface morphology changes and asperity structure evolution using scanning electron microscopy (SEM). Quantitative details pertaining to the surface finish including surface roughness $(S_a)$ and volume of inter-asperity ``valleys" ($S_v$) were measured using white light interferometry. Inter-asperity valleys were characterized by the surface heights lying below $10^{\text{th}}$ percentile on the bearing area curve (i.e., the cumulative distribution of surface profile) \cite{hutchings1992tribology}. To ensure that observations and measurements were made at the same surface location during different polishing steps, the sample surface was initially indented with a $2\times2$ mm square grid. The vertices of this grid enabled us to image the same surface location after each interrupted test. To facilitate better observations of the plastic flow patterns at asperity surfaces, the sample was tilted by $70^{\circ}$ in the scanning electron microscope. 

\begin{figure} 
	\includegraphics[width = 0.8\textwidth]{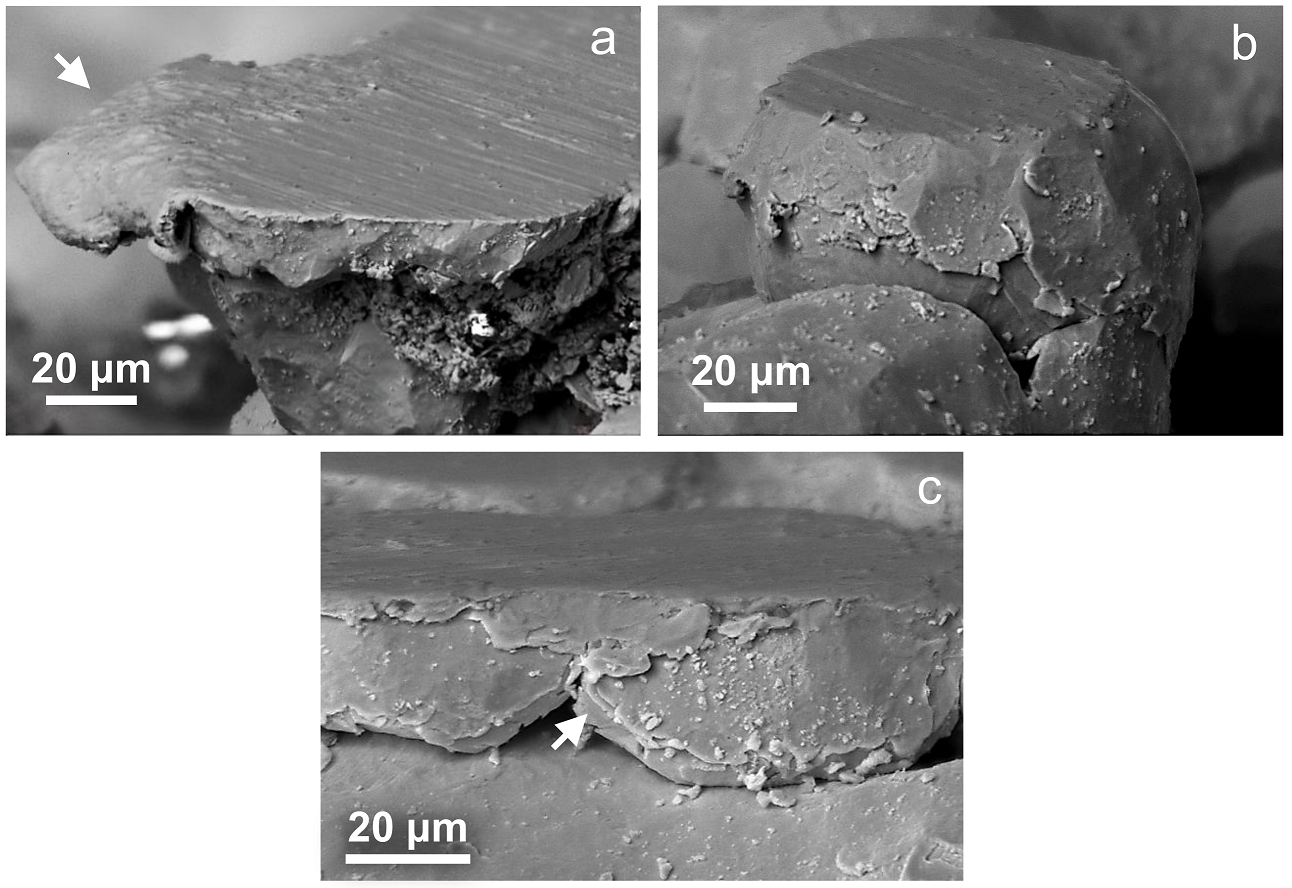}
	\centering
	\caption{SEM images of surface asperities after 90 s of polishing showing (a) shear deformation of the asperity surface and material flow towards the edge of the asperity, and (b) lateral plastic flow and deposition of material on the asperity sides in the form of a thin layer. (c) shows repeated formation of thin material layers as the asperity progressively flattens out on continued polishing.}
	\label{fig:im2} 
\end{figure}

Electron microscopy of the surface asperities enabled us to capture key phenomenological details of the polishing mechanism. Figures~\ref{fig:im2}(b) and \ref{fig:im2}(c) show typical asperity structures after 90 s of polishing. Severe shear of the asperity surface and accumulation of the material towards asperity edges (see at arrow) is evident from Fig.~\ref{fig:im2}(a). This flow pattern is reminiscent of plastic sliding between surfaces oriented at shallow angles, such as in sliding indentation or `machining' under highly negative rake angles \cite{Challen161,komanduri1971some}. Furthermore, the sheared surface material is often seen to flow to the lateral sides of the asperity as thin layers (Fig.~\ref{fig:im2}(b)). Interestingly, the flow is seen to be quite symmetric around the periphery of sheared surface, with deposited material layer showing a molten-like appearance. The sliding direction between the asperity and abrasive particle can be inferred from the sliding marks in Fig.~\ref{fig:im2}(b). This omni-directional flow at the surface, coupled with the observation of rheological flow features at the asperity edges (Fig.~\ref{fig:im2}(a)), suggests fluid-like behavior of the surface plastic flow in polishing.

To explore the possible origin for this flow behavior, we estimated the ``flash'' temperature at the asperity--abrasive sliding contacts using the circular moving heat source model \cite{carslaw1959conduction}, where the abrasive particle was treated as a semi-infinite moving body over which a stationary heat source acts. The heat source intensity was taken as the heat dissipation due to plastic shearing of the asperity at the sliding asperity--abrasive contact. Details of the flash temperature calculations are presented in the Supplemental Material. The analysis showed
that the temperature rise at the sliding contacts monotonically increases with the circular
contact area. The calculated temperatures for $\sim30$\% of the sliding contacts were above 700 K. While these temperatures are well below the melting temperature ($T_m=$  1925 K) of Ti-6Al-4V, they are in the typical dynamic recrystallization temperature range ($700-900$ K) for this alloy where significant flow softening occurs \cite{liao1998adiabatic}. At such temperatures, rate-dependent viscous plastic flow is not uncommon in metals \cite{ashby1982deformation}. Similar fluid-like flow phenomenon in metals have been also noted previously in other sliding configurations \cite{sundaram2012mesoscale,trent2000metal} and shear bands \cite{sagapuram2016geometric,healy2015shear,spaepen2006metallic}.

Our observations of the polished surfaces provide evidence for the surface flow theory in that the surface smoothening is mediated by material redistribution more so than material removal. Figure~\ref{fig:im2}(c) shows the progression of the plastic flow at the asperity surface on continued polishing (beyond 90 s). Apparently, the repeated shearing at the asperity surface upon encountering a sliding abrasive results in stacking of multiple thin layers on the lateral sides of the asperity (see at arrow). In effect, this results in a radial increase in the flattened area of the asperity.

\begin{figure}
	\includegraphics[width = 0.8\textwidth]{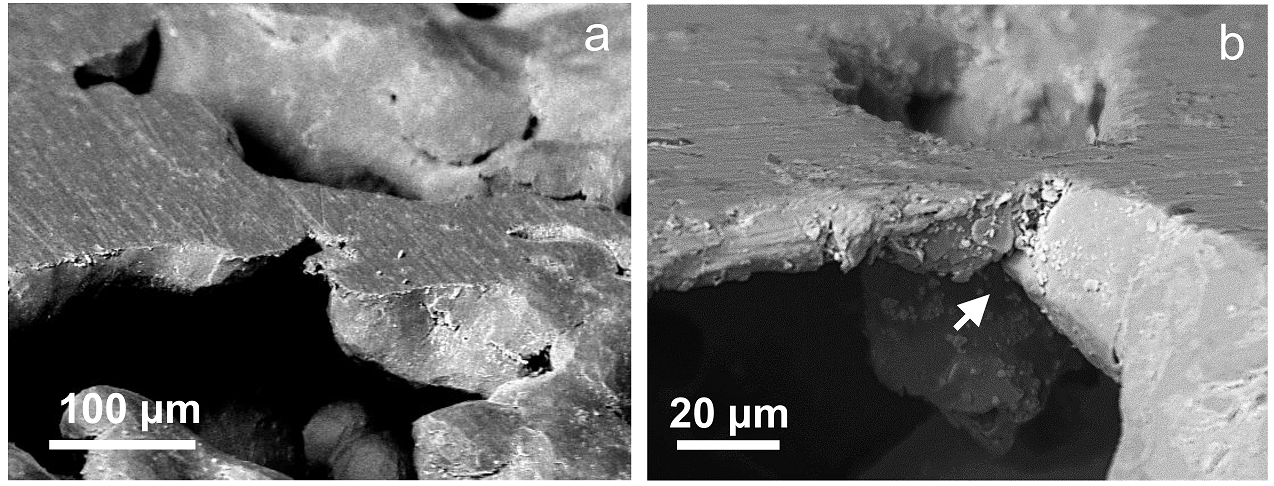}
	\centering
	\caption{Surface smoothening during later stages of polishing by bridging between neighboring asperities: (a) an SEM image showing the interconnection of flat (``smooth") regions surrounded by unfilled depressions; (b) a high-magnification image of the bridge (arrow) that has formed between neighboring asperities.}
	\label{fig:im3}
\end{figure}

\begin{figure}
	\includegraphics[width = 0.8\textwidth]{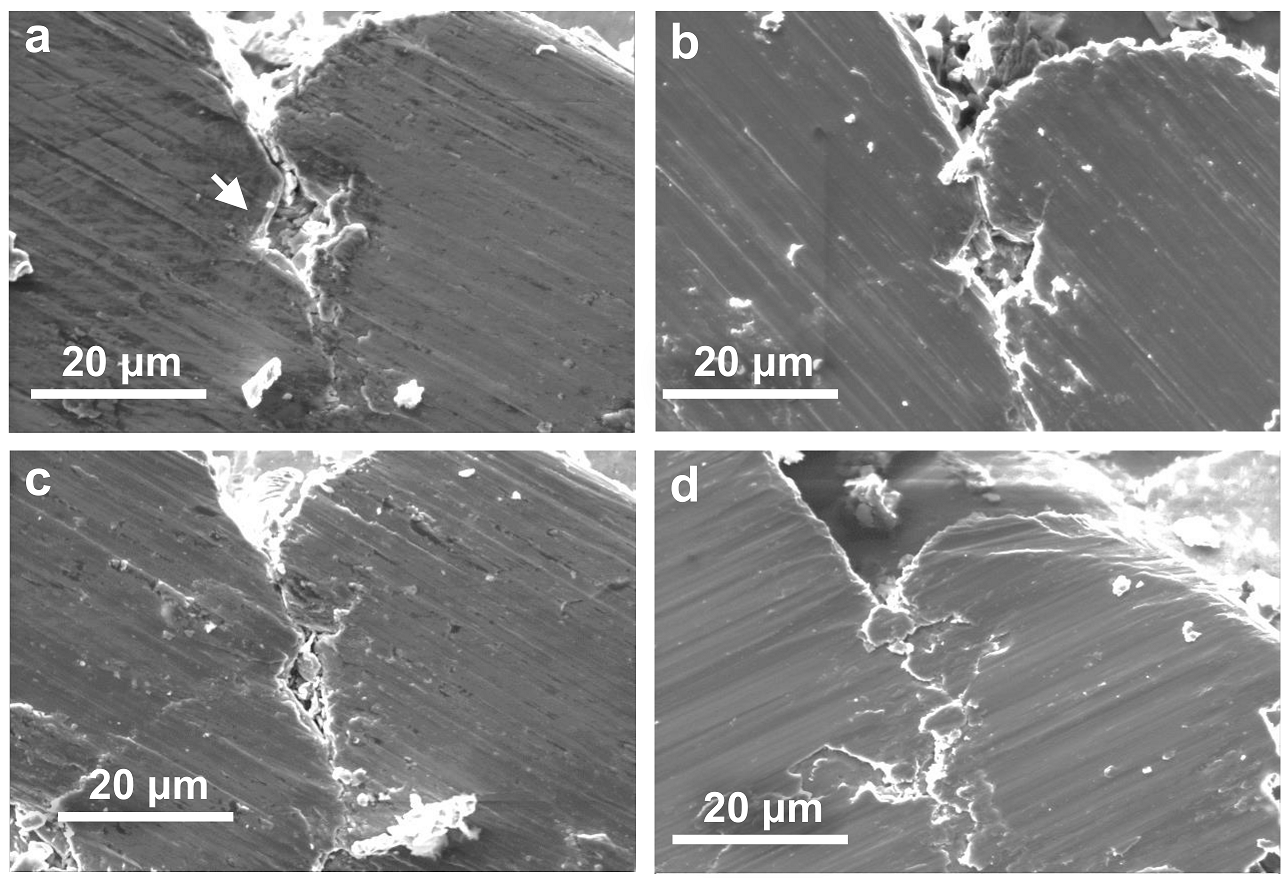}
	\centering
	\caption{SEM observations of a surface depression (indicated by arrow, $\sim10$ $\mu$m in size) showing its temporal evolution under polishing. Images in (a)--(d) are taken at 90 s time interval ($t = 180-450$ s). The depression is gradually filled up as a result of material flow from the surface.} 
	\label{fig:im4} 
\end{figure}

\begin{sidewaysfigure}
	\includegraphics[width=1\textwidth]{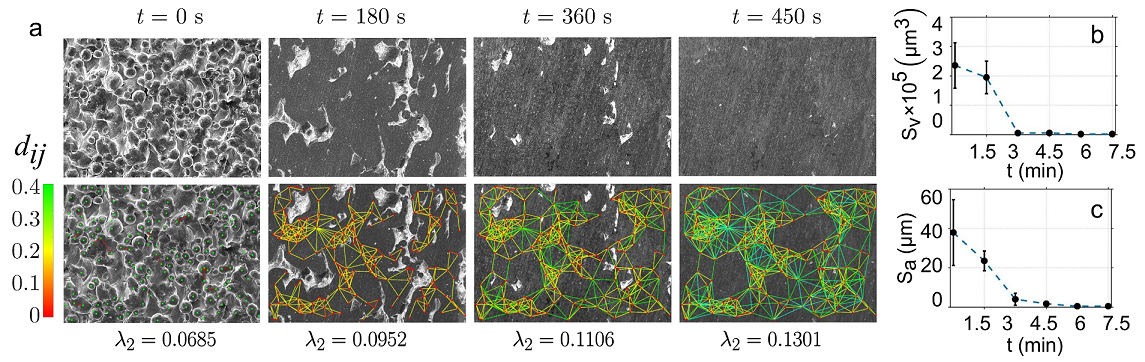}
	\caption{Surface morphology evolution during polishing: (a) SEM images (top row) and corresponding network representation (bottom row) showing the gradual asperity bridging process. The corresponding Fiedler number, $\lambda_2$ (which quantifies the extent of connectivity among the asperities) is also given. The colormap represents the radial distance $(d_{ij})$ of asperity centers, and shows bridging of farther asperities during the later stages of polishing. The plots in (b) and (c) show the temporal evolution of the volume of the voids $(S_{v})$ and the surface roughness $(S_a)$ during polishing.}
	\label{fig:im5}
\end{sidewaysfigure}

Figure~\ref{fig:im3} illustrates the surface morphology characteristics at 180 s. As seen from Fig.~\ref{fig:im3}(a), individual asperity surfaces are unresolvable by this stage, and the surface can be described as an interconnected network of flat islands. Interspersed among these regions are the unfilled depressions. A closer inspection of the flattened regions reveals that their formation is mediated by bridging of the smeared surface material between the neighboring asperities, as shown in Fig.~\ref{fig:im3}(b) (see at arrow). Indeed, this ``welding'' between the asperities may be expected given the occurrence of severe plastic flow and temperatures at the asperity surfaces. In our experiments, this bridging phenomenon was noted only when the distance between the edges of two neighboring asperities was below a critical value of $\sim 30$ $\mu$m. For asperities separated by larger distances, lateral flow of the material (Fig.~\ref{fig:im2}) was seen to continue until the effective distance between the asperities approached the critical value. Continued polishing causes complete bridging of  individual asperities, resulting in a nominally smooth surface (for example, see top row in Fig.~\ref{fig:im5}(a)). The elimination of microscale depressions during final stages of polishing again seems to occur as a result of material flow from neighboring flat regions. A series of SEM images taken at successions of 90 s, and showing the closure of a surface depression, is presented in Fig.~\ref{fig:im4}. An important consequence of repeated plastic flow at the surface is the microstructure refinement at the surface and associated  increase in the strength. Indeed, hardness measurements (Vickers indentation, load $500$ g) showed the surface to be characterized by a higher hardness (375 kg/mm$^2$) compared to the base material (350 kg/mm$^2$).

Figure~\ref{fig:im5} summarizes the surface morphology evolution during the entire duration of polishing. The micrographs show the surface flow and bridging among the asperities (Fig.~\ref{fig:im5}(a), top row), together with the gradual reduction of the volume of inter-asperity valleys (light regions). This results in a strongly connected network of flat areas (dark regions) that eventually evolve to form a uniformly smooth surface (with average roughness, $S_a \sim 30$ nm). The bridging process can in fact be treated as an evolving random graph $G = (V,E)$, where the nodes $(V)$ denote the asperities, and the edges $(E)$ are the probabilities $p(i,j)$ for a bridge to exist between nodes $i$ and $j$ $\forall i,j\in V$ (see Fig.\ref{fig:im5}(a), bottom row). A spectral graph measure called the Fiedler number, $\lambda_2$ \cite{chung1997spectral}, serves as a natural quantifier to capture the effects of polishing on the surface morphology, particularly during the bridging process \cite{rao2015graph}. For example, $\lambda_2=0$ indicates complete absence of bridge formation; in contrast, $\lambda_2= 0.23$ suggests a high degree of bridging where every node is connected to at least six other neighboring nodes (representative of close packing of spherical asperities). Details related to the calculation of $\lambda_2$ are presented in the Supplemental Material. The micrograph patterns as well as the corresponding $\lambda_2$ values presented in Fig.~\ref{fig:im5}(a) suggest that as polishing ensues and the asperity diameters grow, the propensity of neighboring asperities to bridge (i.e., $p(i,j)$) progressively increases. Quantitatively speaking, the initial value of $\lambda_2=0.068$ (see Fig.~\ref{fig:im5}(a), bottom row) indicates little bridging (average number of bridges connecting a node or the ``degree" is $<1$) as reflected in $p(i,j)$ being close to zero between almost all asperities. Specifically, the edges connecting the neighboring nodes are almost absent initially, and low probability edges (red) connect only a sparse set of neighboring nodes. After 450 s of polishing, $\lambda_2$ increases to 0.130, suggesting a higher degree of bridging among all neighboring asperities (degree $\geq4$), and high $p(i,j)$ values. 

The corresponding temporal evolution of $S_{v}$ and $S_a$, captured using surface interferometry, are given in  Figs.~\ref{fig:im5}(b) and \ref{fig:im5}(c), respectively. While both $S_v$ and $S_a$ decrease monotonically with time, $S_v$ drops sharply from $\sim2\times10^5$ $\mu$m\textsuperscript{3} to $2.1 \times10^3$ $\mu$m\textsuperscript{3} between 90 s and 180 s (Fig.~\ref{fig:im5}(b)). This corresponds to the time interval where bridging of the asperities is predominant (see Fig.~\ref{fig:im3}). Unlike $S_v$, the $S_a$ continues to decrease even after 180 s, likely because of surface smoothening via reduction in microscale surface depressions during the final stages of polishing (see Fig.~\ref{fig:im4}).

While this study has focused on a Ti-based alloy system, the current findings are likely to be more generic to polishing of a range of other material systems. In fact, surface flow profiles at the asperities similar to that in Figs.~\ref{fig:im2} and \ref{fig:im3} were also observed during polishing of tantalum oxide (Ta$_2$O$_5$; see Fig.~S4 in the Supplemental Material). These observations in oxide materials, while at first surprising given their inherent brittle behavior, can be explained by the high asperity--abrasive contact pressures that typically exceed the workpiece material's hardness. These high contact pressures can in turn promote plastic flow even in highly brittle materials \cite{bridgman1952studies,marsh1964plastic}. Additionally, the asperity--abrasive contact temperature calculations for polishing of Ta{$_2$}O{$_5$} showed that the flash temperatures can be a significant fraction ($\sim0.4T_m$) of its melting temperature, which could potentially enhance the propensity for viscous-type flow at asperity surfaces. 

In closing, this letter presents direct experimental evidence for the surface flow mechanism of polishing, and reports new phenomenological observations pertaining to plastic flow aspects in smoothening of rough surfaces. These involve material flow from the asperity contact surfaces to the lateral sides in the form of thin viscous layers, bridging of neighboring asperities, and eventual filling-up of the small surface depressions by material flow from the smooth surface regions. While these observations are consistent with the Beilby--Bowden's material redistribution theory of polishing, several important distinctions are in order. First, no evidence for surface melting or amorphization  was noted in contrast to the original hypotheses \cite{beilby1921aggregation,bowden1937physical}, although the microscopy observations of the surface flow profiles, together with the temperature calculations of the asperity--abrasive sliding contacts, strongly suggest the occurrence of viscous flow. Second, as demonstrated in Fig.~\ref{fig:im2}, the material redistribution is facilitated by the material flow as thin layers that make self-contact with the asperity sides. This is again at variance with the original ideas where the surface valleys are believed to be filled purely via mechanical deformation (as in compression or indentation plastic flows) of the asperities. Lastly, bridging among asperities is seen to be an important mechanism by which neighboring asperities merge to form a smooth surface network. This has not been accounted for in any of the prior studies. Besides polishing, our observations are also of relevance to a range of other engineering and physical systems where micro-scale asperity contacts, characterized by high pressures, are of intrinsic interest, e.g., tribological systems, erosion and earthquakes. The well-known observations of the folded-layer structures in metamorphic rocks \cite{hopgood1999general}, which bear striking resemblance to the thin-layer stacking profiles in Fig.~\ref{fig:im2}(c), alludes to the possibility of similar viscous flow phenomena playing a role also on a much larger scale in geophysical formations.

\textbf{Acknowledgments}: The authors would sincerely like to acknowledge Dr. Alex Fang, Texas A\&M University, for providing access to the lapping machine and the National Science Foundation (CMMI- 1538501) for their kind support of this research. 
\renewcommand\bibname{REFERENCES}
\bibliography{references_arXiv_submission_main}
\clearpage
\renewcommand{\thesection}{}
\makeatletter
\def\@seccntformat#1{\csname #1ignore\expandafter\endcsname\csname the#1\endcsname\quad}
\let\sectionignore\@gobbletwo
\let\latex@numberline\numberline 
\def\numberline#1{\if\relax#1\relax\else\latex@numberline{#1}\fi}
\makeatother

\makeatletter

\renewcommand\thefigure{S\arabic{figure}}    
\makeatother
\setcounter{figure}{0}    
\begin{center}
	\textbf{Supplemental Material}
\end{center}

\begin{center}
	\begin{tabular}{ p{5em}  p{18em} }
		\multicolumn{2}{l}{\raggedleft{\textbf{Nomenclature}}} \\
		\hline
		$z$ & asperity heights from the reference plane\\ 
		$a$ & contact radius of the asperity--abrasive contact area\\
		$R$ & asperity radius\\ 
		$\Delta T_{max}$ & maximum temperature rise at the asperity--abrasive interface\\
		$H$ & workpiece surface hardness\\ 
		$V$ & polishing speed\\
		$q$ & total heat flux at the asperity--abrasive interface \\
		$q_1,q_2$ & heat flux at the asperity and abrasive surfaces, respectively\\
		$P_{e1}, P_{e2}$ & Peclet number for the asperity and abrasive body\\
		$\mu$ & coefficient of friction at the asperity--abrasive interface\\
		$k_1,k_2$ & thermal conductivity of asperity and abrasive, respectively\\
		$K_2$ & thermal diffusivity of the abrasive\\
		$\rho_2$ & density of abrasive\\
		$C_2$ & specific heat of abrasive\\
	\end{tabular}
\end{center}

\begin{table}[h]
	\caption{Properties of abrasive and workpiece (asperity) materials.}
	\centering
	\begin{tabular}{ M{8em} M{10em}  M{7em}}	
		\hline
		Material & Thermal conductivity (W/m/K) & Hardness (GPa) \\ 
		\hline 
		Ti-6Al-4V   & 7.2 -- 11.2 [S1] & 3.5 -- 3.75 [S1]\\ 		 
		$\text{Ta}_2\text{O}_5$   & 0.9 -- 4 [S2]  & 1.46 -- 4.21 [S3] \\ 			
		SiC & ~60 [S4]  & 25 [S5] \\ 		\hline 
	\end{tabular}
\end{table}

\begin{center}
	{\textbf{S1: Electron beam melting process parameters}}
\end{center}

Ti-6Al-4V cylindrical disks ($\diameter$50 mm and 7 mm thickness) were prepared using an ARCAM electron beam melting machine operating at a vacuum of $\sim2$ Pa and accelerating voltage of $\sim60$ kV. The process involved raking a $50$ $\mu$m layer of Ti-6Al-4V powder of average $\diameter72$ $\mu$m (see Fig.~S3) for the distribution of radius of Ti-6Al-4V particles) using a focused beam of 3 mA, scanning at a speed of 10 m/s. The resulting surface consists of granular Ti-6Al-4V particles with a unique spherical asperity structure. Such a controlled asperity structure is ideal for systematic investigation of the surface flow behavior during polishing.

\begin{center} 
	{\textbf{S2: Calculation of flash temperatures at the asperity--abrasive contacts}}
\end{center}

\begin{figure}[h]
	\includegraphics[width=0.75\textwidth]{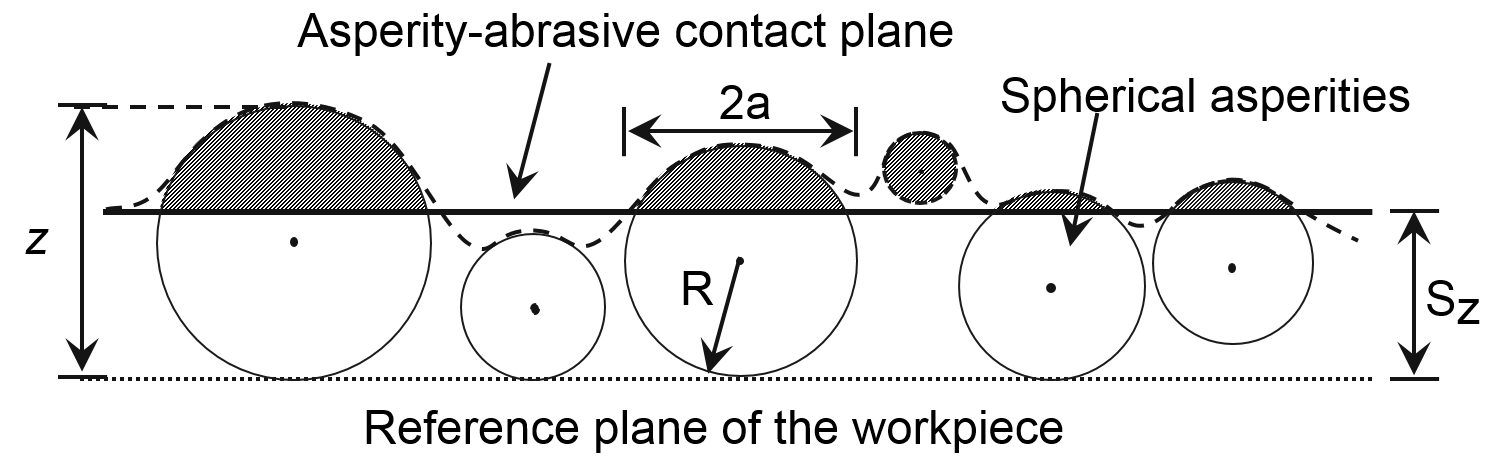}
	\centering
	\caption{Schematic showing contact (solid line) between the workpiece surface consisting of spherical asperities and the polishing pad at a distance $S_z$ (average asperity heights) from the workpiece reference plane (dotted line). Here, the asperity height, $z$, is measured with respect to the workpiece reference plane.}
	\label{fig:suppim1}	
\end{figure}

\begin{figure}
	\includegraphics[width=0.75\textwidth]{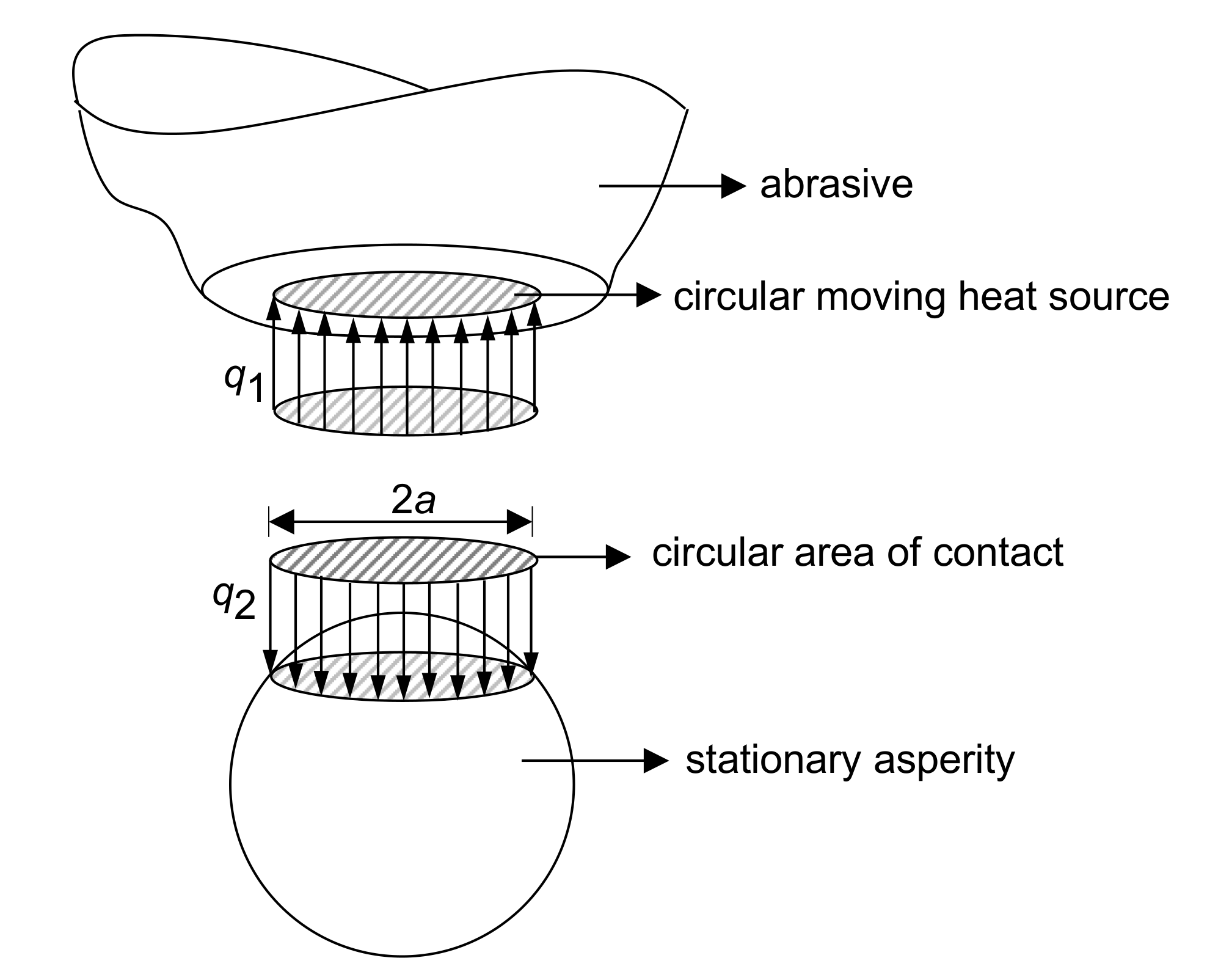}
	\centering
	\caption{Moving circular heat source model for the contact between asperity and abrasive to calculate the temperature rise during polishing. Here, the abrasive is considered as the semi-infinite moving heat source and the asperity acts as a stationary heat source.}
	\label{fig:suppim2}	
\end{figure}
For a given asperity height $(z)$ distribution, only the asperities for which $z>S_z$ and $z\leq S_z+2R$ are involved in the polishing process, as schematically shown in Fig.~S1. Here, the asperity height, $z$, is measured with respect to the workpiece reference plane (dotted line in the schematic in Fig.~S1). We assume that the clearance between the workpiece reference plane and the polishing pad (solid line) is equal to the average surface asperity heights, $S_z$, of the workpiece. The diameter of asperity--abrasive contact ($2a$) can then be calculated for a given value of $S_z$, asperity radius ($R$) and height ($z$) distribution.\\

Given the radius of contact, we calculate flash temperature by treating the contact as a moving circular heat source (Fig.~S2). The heat source intensity is taken as the heat dissipation due to plastic shearing of the metal asperity at the sliding interface. The heat partition between the asperity and the abrasive particle is determined by setting equal the maximum (quasi-steady state) temperatures of the asperity and abrasive particle within the contact, according to Blok's postulate [S6]. Here, we treat the abrasive as a semi-infinite moving body (with velocity $V$) over which a stationary heat source (with uniform heat flux) acts. The steady state flash temperature occurring at the contact center can accordingly be given by the first order approximation to Jaegar's circular moving heat source model [S7, S8] as: 
\begin{equation}
\Delta T_{max}\big|_{abrasive}=\frac{2q_2a}{k_2\sqrt{(\pi(P_{e2}+1.273))}}
\end{equation}
where, Peclet number, $P_{e2}=Va/2K_2$ and $K_2={k_2}/{\rho_2 C_2} \approx 4\times10^{-5}$ m$^2$/s. For $V=5$ m/s and contact radius $a$, we have $P_{e2}=6.25\times10^5a$. For the asperity (which is treated as a stationary source), we have:
\begin{equation}
\Delta T_{max}\big|_{asperity}=\frac{q_1a}{k_1}
\end{equation}

\begin{figure}	
	\centering
	\includegraphics[width=1\textwidth]{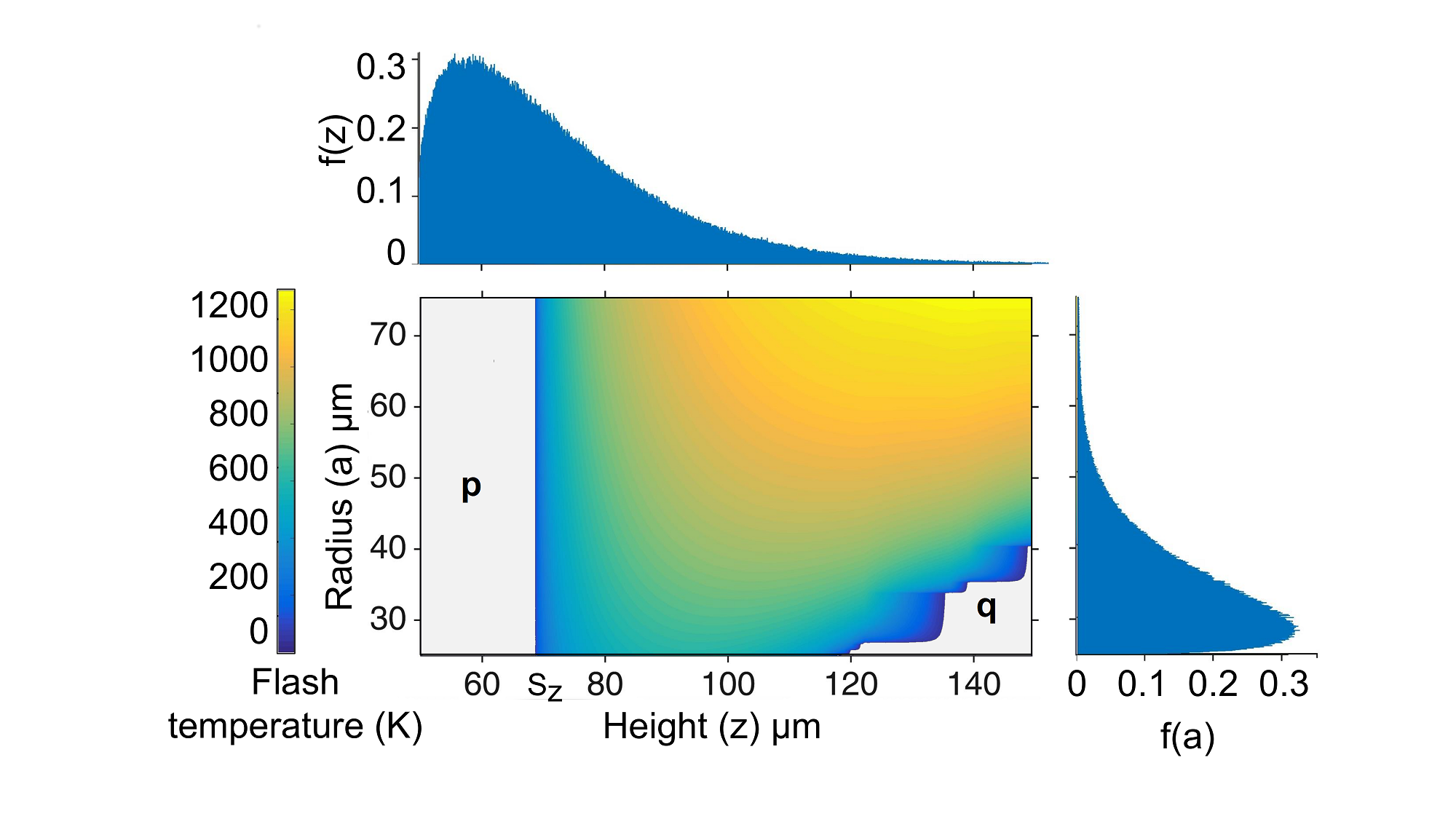}
	
	\caption{(a) Flash temperature map for Ti-6Al-4V as a function of asperity radius and height, both of which follow a truncated Weibull distribution with average at 36 $\mu$m and 64.5 $\mu$m, respectively, and a standard deviation $\sim15$ $\mu$m. $S_z$ corresponds to the average asperity height. }
	\label{fig:suppim3}	
\end{figure}

Assuming adiabatic conditions, where plastic dissipation at the interface is completely converted into heat, the total heat flux, $q$, at the circular contact is given by: 
\begin{equation}
q =q_1+q_2=\mu H V
\end{equation}
By equating the maximum temperatures at the asperity and abrasive surface, we have:  
\begin{equation}
\Delta T_{max}=\frac{\mu HVa}{k_1}\left(1+\frac{k_2}{2k_1}\sqrt{\pi(P_{e2}+1.273)} \right)^{-1}
\end{equation}

We solve for $\Delta T_{max}$ for Ti-6Al-4V using the values in Table 1, and the corresponding flash temperature map as a function of asperity height and radius is shown in Fig.~S3(a). Any asperity for which $z<S_z$ or $z\geq S_z+2R$ would not be involved in the polishing process as it would either make no contact with the abrasive or lie outside the asperity--abrasive contact region (solid line in Fig.~S1). These two cases are marked as ``\textbf{p}" and ``\textbf{q}" in Fig.~S3(a). Elsewhere, we notice that larger values of $R$ and $z$ result in higher flash temperatures. 

{\color{black}While the assumption of abrasive as a semi-infinite plane maybe reasonable during the initial stages of polishing, the configuration is reversed as polishing process progresses. During the intermediate and final stages, polishing maybe represented as individual abrasive particles sliding across a semi-infinite workpiece surface. For this latter configuration, we  assume abrasive particles as sliding conical indenters plastically deforming the workpiece surface. Again for this case, the problem is that of a moving semi-infinite body (workpiece surface) over which stationary heat source (abrasive-workpiece surface contact) acts. The maximum flash temperature rise at the contact in this case is given as:
	\begin{equation}
	\Delta T_{max}=\frac{\mu HVa}{k_2}\left(1+\frac{k_1}{2k_2}\sqrt{\pi(P_{e1}+1.273)} \right)^{-1}
	\end{equation}
	The calculated sliding temperatures for this configuration are slightly larger than those in the earlier configuration where abrasive was taken as a semi-infinite plane (Fig. S2). The difference between temperature estimates for these two configurations is within 20\% (at a contact radius of $\sim$40 $\mu$m) for the contact areas considered here.} In both the configurations, for $\sim30\%$ of the sliding contacts, maximum flash temperatures are above the dynamic recrystallization temperature of the alloy ($\sim700$ K).

Similar calculations for $\text{Ta}_2\text{O}_5$ showed the flash temperature to be in the range of 750~K. In this case, the average radius of the asperity--abrasive contact area was inferred from Fig.~S4 as $\sim15$ $\mu$m. $V$ was taken as 5 m$/$s, as for Ti-6Al-4V polishing. Again the calculated flash temperatures at the asperity--abrasive contacts are high enough, $\sim0.4T_m$, where viscous-like flow may be expected.\\

\begin{figure}
	\includegraphics[width=0.75\textwidth]{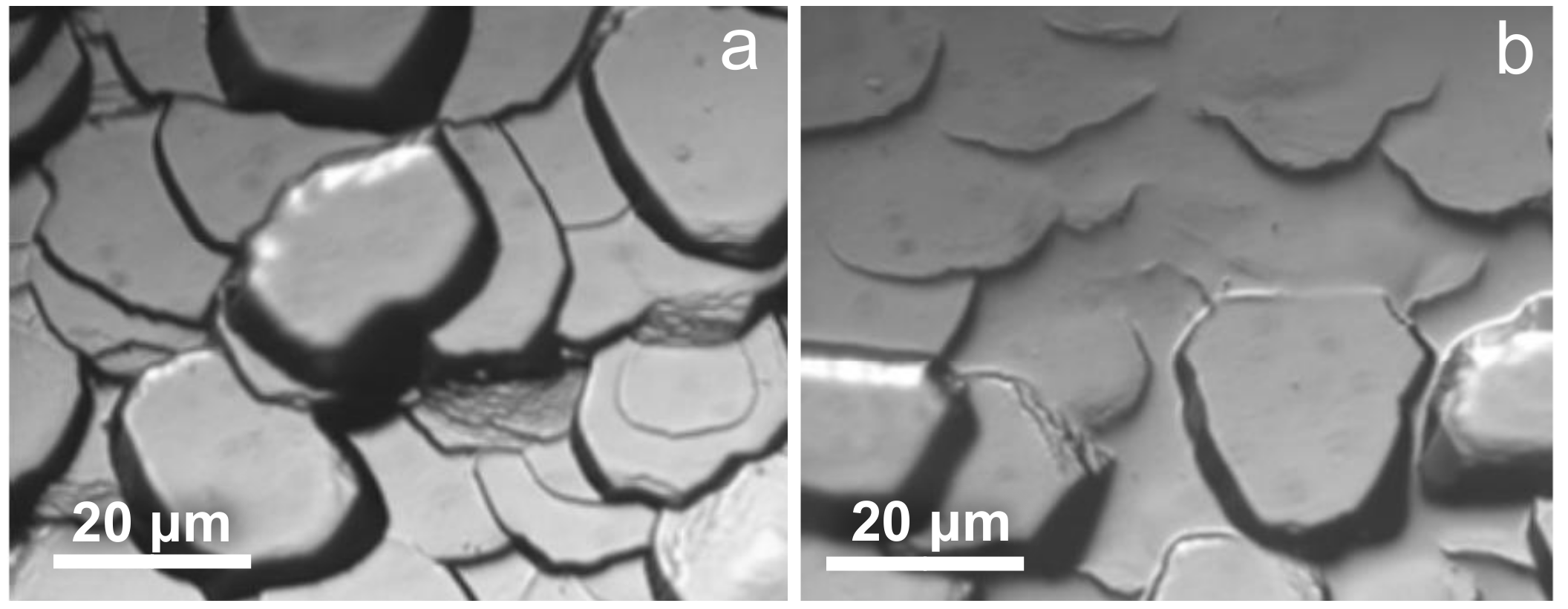}
	\centering
	\caption{Scanning electron micrographs showing surface morphological changes in $\text{Ta}_2\text{O}_5$: (a)~before and (b) after polishing. }
	\label{fig:suppim4}	
\end{figure}

\begin{sidewaysfigure}
	\includegraphics[width=\textwidth]{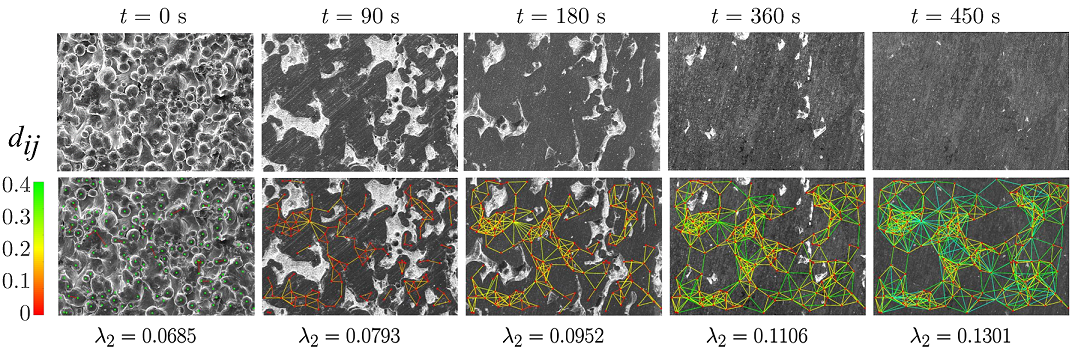}
	\centering
	\caption{Topological evolution of the Fiedler number ($\lambda_2$) as a function of time. Top row shows a series of SEM images taken at 90 s interval. The corresponding asperity network in each of these stages is shown in the bottom row.}
	\label{fig:suppim5}	
\end{sidewaysfigure}
\vspace{2cm}

\begin{center}
	\textbf{S3: Graph representation of topological evolution}
\end{center}

The process of the merger (connectivity) among the asperities is analyzed as an evolving random graph {$G=(V,E)$} whose nodes ($V$) are the asperities, and the edges ($E$) are given by the probability, $p_{ij}$, of the existence of a bridge connecting the asperities $i,j\in V$. The probability of existence of an edge ($p_{ij} \forall i,j\in V$) is inversely proportional to the inter-asperity distance and is calculated using the radial basis function as $p_{ij}=(1+\exp(||V_i-V_j||))^{-1}$. The radial basis function assigns lower $p_{ij}$ to the edges as the physical distance between connecting node increases. We first determine the normalized Laplacian, $\mathcal{L}$, from the graph, $G$ as $\mathcal{L=D}^{-\frac{1}{2}}\times L\times \mathcal{D}^{-\frac{1}{2}}$ where $L$ is the combinatorial Laplacian defined as $L\overset{\Delta}{=}\mathcal{D-S}$. Here, $\mathcal{D}$ is the dianognal matrix representing the degree of each node and is given as $\mathcal{D}=\left( \begin{matrix}
\sum_{j=1}^{N}p_{1j}&\sum_{j=1}^{N}p_{2j}&...&\sum_{j=1}^{N}p_{Nj} \end{matrix}\right) $, and $\mathcal{S}$ being the similarity matrix. It has been established that the second largest eigenvalue of $\mathcal{L}$ captures the algebraic connectivity in the graph, also called the Fiedler number ($\lambda_2$) [S9, S10]. 

The lower bound on $\lambda_2$ is calculated using the geometric embedding of planar graph on a unit sphere as presented in [S11], where each of the nodes are represented by non-overlapping semi-spherical caps of radius $r_i, i\in V$.
For the micrograph in Fig.~S5, $|V|=160$. A strongly connected network of asperities can be assumed as an ideal close packing of uniform spheres such that each node is connected to at most 6 nearest neighbors. Under such conditions it can be shown that $0.23\leq\lambda_2\leq0.3$ holds. The initial value of $\lambda_2=0.068$ (see Fig.~S5, bottom row) indicates that the degree of each node is $<1$. After 450~s of polishing, $\lambda_2$ increases to 0.130 suggesting a minimum degree of 4 among all neighboring asperities. The network structure along with the corresponding $\lambda_2$ values is summarized in Fig.~S5. Additionally, the linear increase in the value of $\lambda_2$ suggests that there are significant topological changes in the surface even during the final stages of polishing process which otherwise are not reflected in the $S_a$ or $S_v$ measurements (see Fig. 5 in the main text).

\begin{center}
	\begin{tikzpicture}
	\draw [line width=0.20mm] (-1.30,0) -- (1.30,0);
	\draw [line width=0.30mm] (-0.90,0) -- (0.90,0);
	\draw [line width=0.40mm] (-0.5,0) -- (0.5,0);
	\end{tikzpicture}
\end{center}

\small
\begin{enumerate}[label={[}{S}{\arabic*}{]}]
	\item G. Welsch, R. Boyer, and E. Collings,\textit{ Materials Properties Handbook: Titanium Alloys} (ASM	International, Materials Park, OH, 1993).
	\item C. D. Landon, R. H. Wilke, M. T. Brumbach, G. L. Brennecka, M. Blea-Kirby, J. F. Ihlefeld,
	M. J. Marinella, and T. E. Beechem, Applied Physics Letters \textbf{107}, 023108 (2015).
	\item O. Shcherbina, M. Palatnikov, and V. Efremov, Inorganic Materials \textbf{48}, 433 (2012).
	\item Q. Liu, H. Luo, L. Wang, and S. Shen, Journal of Physics D: Applied Physics (2016).
	\item Y. Ahn, S. Chandrasekar, and T. N. Farris, Journal of Tribology 119, 163 (1997).
	\item H. Blok, in \textit{Proceedings of the general discussion on lubrication and lubricants}, Vol. 2 (London:
	IMechE, 1937) pp. 222-235.
	\item H. S. Carslaw and J. C. Jaeger, \textit{Conduction of Heat in Solids} (Clarendon Press, Oxford, 1959).
	\item X. Tian and F. E. Kennedy, Journal of Tribology \textbf{116}, 167 (1994).
	\item F. R. K. Chung, \textit{Spectral Graph Theory}, Vol. 92 (American Mathematical Society, RI, 1997).
	\item P. K. Rao, O. F. Beyca, Z. Kong, S. T. Bukkapatnam, K. E. Case, and R. Komanduri, IIE Transactions \textbf{47}, 1088 (2015).
	\item D. A. Spielman and S. H. Teng, Linear Algebra and its Applications \textbf{421}, 284 (2007).
\end{enumerate}

\end{document}